\documentclass[cup6b]{cupbook}

\usepackage{epsf}
\usepackage{hyperref}

\renewcommand{\baselinestretch}{0.99}

\def\ie{{\em i.e.}}

\def\note #1]{{\bf #1]}}

\def\note #1]{{\bf #1]}}

\def\sp{{Silly~Putty\textregistered} material}  

\begin{document}

\author{PETER T. WILLIAMS\\
Los Alamos National Laboratory,  Los Alamos, NM, USA
}

\chapter{Turbulent Magnetohydrodynamic Jet Collimation\\ and Thermal Driving}
\renewcommand{\baselinestretch}{0.95}
{\it 
We have argued that magnetohydrodynamic (MHD) turbulence in an accretion disk
naturally produces hoop-stresses, and that
in a geometrically-thick flow these stresses could both drive and
collimate an outflow. We based this argument on an analogy of turbulent
MHD fluids to viscoelastic fluids, in which azimuthal shear flow creates
hoop-stresses that cause a variety of flow phenomena, including the
Weissenberg effect in which a fluid climbs a spinning rod. 

One of the more important differences between the Weissenberg effect and
astrophysical jets is the source of power. In our previous analysis, we
only considered the power due to the spin-down torque on the central
object, and thus found that we could only drive an outflow if the central
object were maximally rotating. Here we take into account the energy
that is liberated by the accreting matter, and describe a scenario in
which this energy couples to the outflow to create a thermodynamic
engine. 
}
\renewcommand{\baselinestretch}{0.99}
\section{Introduction}

We wish to discuss here
in simple language some of our ideas regarding jet collimation and acceleration.
In this paper, we will concentrate on the basic intuitive notions 
rather than the mathematics, which we have discussed in print
elsewhere (see references below).


\section{Review: Turbulence Models and Jets}
We have argued (Williams 2001, 2003b; see also Ogilvie 2001, 2003)
 that the stress due to magnetohydrodynamic (MHD) 
turbulence in ionized accretion disks --- such as, but not limited to, the
turbulence driven by the magnetorotational
instability (MRI) --- behaves more like the stress in a viscoelastic fluid than
the stress in a viscous fluid. 

Viscoelastic fluids are a broad category of non-Newtonian fluids. 
Biological liquids with high
nucleic acid or protein polymer concentrations, such as mucus, are one example. 
American popular culture provides the more pleasant example of \sp,
originally made by polymerizing silicone oil with boric acid.

These fluids are
``goopy'' and ``stretchy''; the stress in them does not relax quickly, putting them
somewhere between Newtonian fluids and elastic solids. The tangled polymers take time
to rearrange themselves when the fluid suffers a distortion. 
Now let us replace these
tangled polymers with a tangled magnetic field in a conducting fluid. 
Since we are taught that
magnetic fields in ideal MHD are like elastic strings under tension, 
it seems natural to think of a tangled
mess of magnetic fields as a goopy, stretchy fluid, as well. 
This is by no means a perfect analogy, but we argue that it is a better analogy than
that provided by purely {\em viscous} fluids.

Deformation of a viscoelastic fluid will typically create different normal stresses 
in different directions. 
This is why it is possible to wrap \sp\ 
around one's finger while temporarily maintaining the material under linear tension.
Note that in an orthogonal coordinate system,
normal stresses are the stress components along
the diagonal of the stress tensor; shear stresses are off-diagonal.
These fluids also have differing normal stresses in azimuthal shear flow,
exhibiting tension in the azimuthal direction. This is a hoop-stress.

Such stress is responsible for various
effects not seen in ordinary Newtonian fluids. 
One such effect we have discussed previously is that a spinning sphere immersed in
a viscoelastic fluid may cause a meridional circulation --- \ie, a poloidal flow --- 
in which fluid is pulled in
along the equator and pushed out at the poles. This is opposite to the meridional
flow in a purely viscous fluid, and it is closely related to
the Weissenberg
effect, in which a fluid climbs a spinning rod.

By the analogy we have drawn between MHD turbulence and viscoelastic fluids, 
we expect that the turbulence in accretion disks 
naturally creates hoop-stresses as well. 
In fact, these hoop-stresses can be seen in simulations
of the MRI (Hawley, Gammie \& Balbus 1995), as pointed out by Williams (2003a). 
Hawley et al. did not discuss these
hoop-stresses, but one has only to look at their tables for the various components
of the stress tensors to see them. In shearing-sheet simulations 
of disk turbulence, such as they performed,
a hoop-stress appears as a stress in the direction of shear. 
These hoop-stresses are of the right order of magnitude to collimate jets.
This is related to, but distinct from, the notion put forth (Akiyama \& Wheeler 2002, Akiyama et al. 2003)
that the MRI saturates to create an azimuthal field that collimates jets:
The hoop-stresses in our picture are created entirely by a tangled field. 



All of this has been discussed previously by us. Here we only briefly comment on the source 
of energy for an outflow. This is related in a simple way to the flow topology, which
we discuss below.


\section{Flow Topology and Jet Thermodynamics}


Let us first sketch the basic arrangement of the streamlines of 
flow for the laboratory phenomenon mentioned above. Figure~1.1-A shows the meridional
flow of a Newtonian
fluid at low Reynolds number. 
The azimuthal Stokes flow is not shown, but it
corresponds simply to corotation at the surface of the sphere, 
gradually dropping to zero at infinity.
This Stokes flow is driven by the viscous torque or couple on the sphere
and the outwards transport of angular momentum to the surrounding medium. This outward transport of
energy is directly analogous to the outward ``viscous'' transport of energy in radiatively efficient thin-disk theory
that is responsible for the infamous factor of three discrepancy between the local dissipation and the thermal emission
(see Frank, King \& Raine 2002, p.~86--87).
The azimuthal flow
drives the secondary meridional flow by the centrifugal (``inertial'') forces.

The ratio of the stress relaxation time of the fluid to a representative timescale
for the flow (such as the rotation timescale in this case) is variously called the
Weissenberg number or the Deborah number.  If we keep the
Reynolds number low but gradually increase the Weissenberg--Deborah number,
the secondary (meridional) flow reverses to the flow qualitatively sketched
in fig~1.1-B, as the elastic hoop-stresses begin to dominate the inertial forces.

The exact flow pattern depends not just on the Reynolds number and the Weissenberg number, but
on the full rheological properties of the material.
For example, the flow may become more topologically complex, with the introduction of a flow
separatrix as shown in figure~1.1--C. Broadly speaking, however, the viscoelastic fluid has stresses
that tend to act opposite to the inertial stresses, as shown.


For the hypothetical analogous 
astrophysical phenomenon, we assume that the flow
is steady and axisymmetric as it is in the laboratory case. 
Of course, if the astrophysical flow is turbulent, as we
are assuming, then streamlines must be interpreted as some type of average over the
turbulent fluctuations; we assume here that it is possible to define such an average.

The source of power
for this flow is the torque of the sphere on the surrounding medium, 
so that for the analogous astrophysical problem,
we found that, roughly, we could only power a jet strong enough to reach escape velocity if the central object were spinning
near the centrifugal limit. This is because the natural scaling for the jet launch speed in this scenario is the surface rotation speed of the star.
Of course, in the astrophysical case, there is also an inward advective transport of angular momentum, as well as gravity and compressibility,
but we still feel this analogy may be informative.

\begin{figure}
\begin{center}
\leavevmode\epsfxsize=10cm \epsfbox{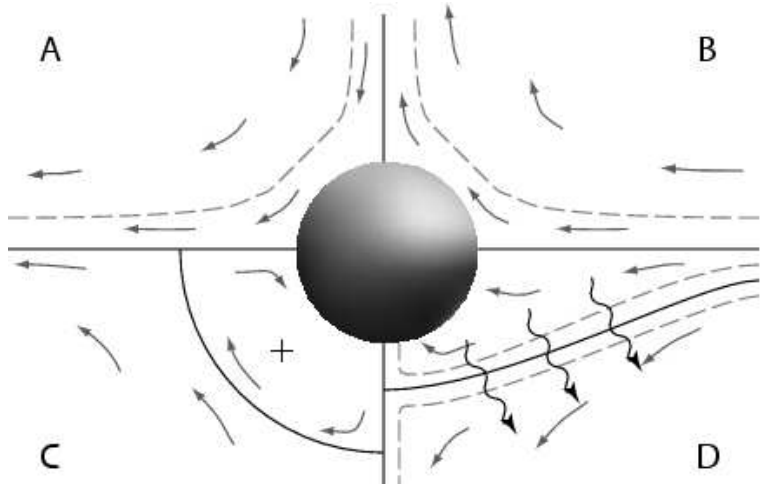}
\end{center}
\caption{
A rough sketch of the meridional flow in three laboratory experiments (A-C), and the hypothetical flow in one
astrophysical case (D).
In the latter case, the sphere is somewhat fancifully representative of a generic
central object. Flow is qualitatively indicated with curved arrows and fiducial streamlines (dashed lines).
Separatrices are shown as solid lines. All flows are axisymmetric and vertically mirror-symmetric. Squiggles represent
flow of thermal and mechanical energy; ``+'' denotes an {\tt o}-point.
}
\label{fig:flow}
\end{figure}

The
mass outflow in an astrophysical
 jet does not need to
equal the mass inflow in the disk, unlike the laboratory phenomenon described above in which the central sphere
acts as neither a source nor sink of material.
Let us follow a fluid element in the accretion flow. This fluid element either ends up being
accreted onto the central object or it does not. It makes sense then to introduce a streamline that separates the flow of material
that ends up being accreted from the material that does not, {\em i.e.} a separatrix. Although in principle the topology of this separatrix could be quite
complex, let us imagine it to be quite simple, as we have drawn it in figure~1.1-D.

As has been long appreciated, a Keplerian thin disk
accreting onto a stationary central sphere dissipates fully {\em half} of its energy in the boundary layer where the aziumthal
velocity is slowed from Keplerian to zero.
 Even in the more general
case where the flow is not Keplerian (nor geometrically thin), and the central object is rotating as we have assumed previously, a relatively very large amount
of energy is released in the central few ``stellar radii'' of the accretion flow.
If some of the energy released by material that ultimately accretes onto the central star or object
could be transmitted to the material that does not accrete, we might have a very effective way of powering a jet. This in itself is not a novel idea, but
we discuss this notion within the context of our previous work.
The common term ``central engine'' would be very appropriate for this driving mechanism; just as in the original steam engine, a
working fluid (\ie~the material that ultimately ends up in the jet) would be compressed, 
heated by a furnace (which in this case is the central accretion region, powered by the
release of gravitational binding energy of the matter that is ultimately accreted), and allowed to expand again. 
This would represent a jet driven in part by the reservoir of angular momentum in the central object, and in part
by the energy of accretion. 

The contribution of energy in the form of heat and $P\,dV$ work 
to the outflow may be visualized as an open virtual thermodynamic engine cycle, as shown in
figure~1.2. This picture does not take into account the mechanical spindown energy that we discussed previously, however.


%

\begin{figure}
\begin{center}
\leavevmode\epsfxsize=10cm \epsfbox{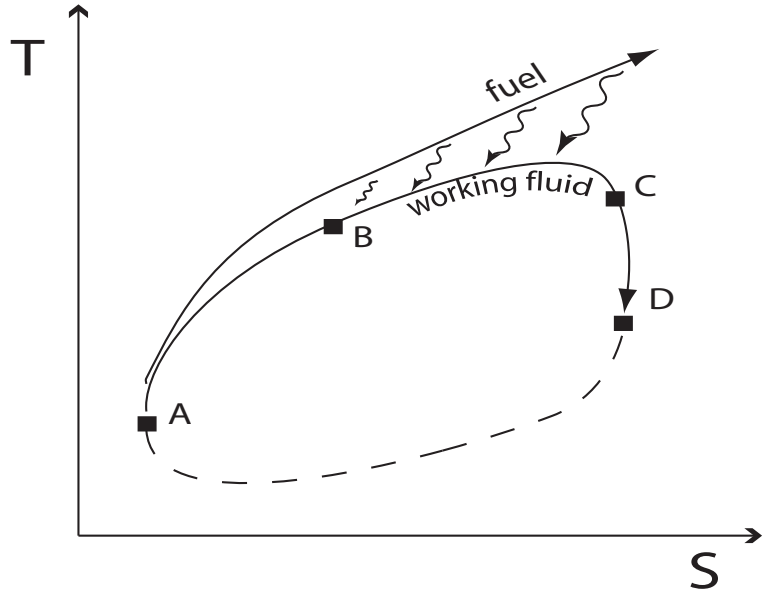}
\end{center}
\caption{
The thermal cycle of the ``central engine.'' 
The first leg ($A \rightarrow B$) is accretion in a relatively thin disk.
Much more energy is transferred to the working fluid in the second leg ($B \rightarrow C$), 
from the fuel that is accreted onto the central
object. The third leg ($C \rightarrow D$) is expansion into an outflow. The fourth leg -- reincorporation into the ISM followed by accretion
again -- is presented as a conceptual aid ($D \rightarrow A$). 
$T$ is temperature and $S$ is entropy.
}
\label{fig:cycle}
\end{figure}

\section{Conclusion}

We argue that independent of any {\em dynamo} process (except perhaps in the loosest sense of the word),
 collimating forces are easily created by
the same MHD turbulence responsible for angular momentum transport in disks, so that
angular momentum transport and outflow collimation may be different sides of the same 
coin. 
The argument may be made {\em a forteriori} in the case of the MRI because it is an MHD rather than a hydro instability,
but the turbulent MHD collimation will exist in either case. We have simply addressed here the consideration that
the spin-down stress is not the only source of power.
Secondly, we wish to make quite explicit the notion of a central {\em engine} as a 
thermodynamic engine, in which there is a fuel and a working fluid. These materials
are of the same substance, the difference between the two being whether the material
is ultimately accreted or not.



Finally, we note here that simulations of nonradiative accretion flows seem to show
that the MRI collimates jets (Hawley \& Balbus, 2002). Note also though that this collimation was claimed
to be due to magnetic {\em pressure}. The collimation we have described here and in our
previous work, however, is due
to the so-called curvature term --- \ie, magnetic tension. 




\bigskip\noindent
{\it Acknowledgements} I am grateful to Paul Bradley, my advisor at
Los Alamos, for giving me the time to present this work, and to 
Craig Wheeler for his constant encouragement. 

\renewcommand{\baselinestretch}{0.97}


\end{document}